\documentclass[letterpaper, 10 pt, conference]{ieeeconf}
\IEEEoverridecommandlockouts % This command is only
\usepackage{amssymb,amsmath,graphicx,float}
\usepackage{algorithm, algpseudocode}
\usepackage{booktabs}
\usepackage{wrapfig}
\usepackage{setspace}
\usepackage{graphicx}
\usepackage{epsfig}            
\usepackage[all]{xy}                             
\usepackage{verbatim}
\usepackage{float}
\usepackage{mathrsfs}  
\usepackage{bm}
\usepackage{float}
\usepackage{booktabs}
\usepackage{algorithm, algpseudocode}
\newtheorem{theorem}{Theorem} 

\newtheorem{problem}{Problem}

\newcommand{\tabincell}[2]{\begin{tabular}{@{}#1@{}}#2\end{tabular}}
\newcommand{\norm}[1]{\left\lVert#1\right\rVert}

\usepackage{fixltx2e}
\usepackage{tabularx}
\usepackage{verbatim}
\usepackage{color}
\usepackage{hyperref}
\usepackage{xmpmulti}
\usepackage{transparent}
\usepackage{fancyhdr}

\begin{document}
	
\title{Robust Pandemic Control Synthesis with Formal Specifications: A Case Study on COVID-19 Pandemic}

\author{Zhe~Xu\thanks{Zhe~Xu is with the School for Engineering of Matter, Transport, and Energy, Arizona State University, Tempe, AZ 85287. email: {\tt\small $\{$xzhe1@asu.edu$\}$. }}, Xiaoming Duan\thanks{Xiaoming Duan is with the Oden Institute for
Computational Engineering and Sciences, University of Texas, Austin, Austin, TX 78712. Email: {\tt\small $\{$xiaomingduan.zju@gmail.com$\}$. }} }

\maketitle 

\begin{abstract} 
Pandemics can bring a range of devastating consequences to public health and the world economy. Identifying the most effective control strategies has been the imperative task all around the world. Various public health control strategies have been proposed and tested against pandemic diseases (e.g., COVID-19). We study two specific pandemic control models: the susceptible, exposed, infectious, recovered (SEIR) model with \textit{vaccination} control; and the SEIR model with \textit{shield immunity} control. We express the pandemic control requirement in \textit{metric temporal logic} (MTL) formulas. We then develop an iterative approach for synthesizing the optimal control strategies with MTL specifications. We provide simulation results in two different scenarios for robust control of the COVID-19 pandemic: one for vaccination control, and another for shield immunity control, with the model parameters estimated from data in Lombardy, Italy. The results show that the proposed synthesis approach can generate control inputs such that the time-varying numbers of individuals in each category (e.g., infectious, immune) satisfy the MTL specifications with robustness against initial state and parameter uncertainties. 
\end{abstract}  

% \begin{IEEEkeywords}
% 	COVID-19 pandemic, robust control, vaccination, shield immunity, metric temporal logic
% \end{IEEEkeywords}

\section{Introduction}
% The COVID-19 pandemic \cite{Fauci2020} has caused over 15 million confirmed cases and over 0.6 million deaths globally as of July 23, 2020. Ever since the outbreak of COVID-19, various public health control strategies have been proposed and tested against the coronavirus the pandemic disease \cite{Stewart_magazine_2020}. 
Pandemics can bring a range of devastating consequences to public health and the world economy. Identifying the most effective control strategies has been the imperative task all around the world. Various public health control strategies have been proposed and tested against pandemic diseases (e.g., COVID-19). However, the existing pandemic control synthesis approaches still suffer from several limitations:
(a) The current control
synthesis approaches do not take into account the uncertainties in the states and parameters. (b)
There is a lack of specific and formal \textit{specifications} for the
expected effects and outcomes of the control strategies.

A specification for a biological system describes its desirable behaviors and formalizes its properties. On the population-level, specifications such as ``\textit{the infected people should never exceed one thousand per day within the next 90 days, and the immune people should eventually exceed 6 million after 40 to 60 days}'', which can be expressed as a temporal logic formula $\Box_{[0,~90\textrm{d}]} (InfectedPerDay\le 1000)\wedge\Diamond_{[40\textrm{d},~ 60\textrm{d}]}(Immune\ge6,000,000)$, can be used for the formal synthesis of pandemic control strategies such as vaccination, quarantine, and shield immunity. Such temporal logic formulas have been used as high-level knowledge or specifications in many applications in artificial intelligence \cite{zhe_ijcai2019}, robotics \cite{Verginis2019Icra}, power systems \cite{zhe_control}, etc. On the agent-level, temporal logic formulas can express specifications for people in both indoor and outdoor spaces to obey the pandemic requirements (e.g., social distancing requirements). For example, according to the United States Centers for Disease Control and Prevention (CDC) guidelines, a \textit{close contact} is defined as ``\textit{any individual who was within 6 feet of an infected person for at least 15 minutes starting from 2 days before illness onset until the time the patient is isolated}''. This can be expressed by a temporal logic formula $\Diamond_{[t_1-2\textrm{d}, t_2]}\Box_{[0,15\textrm{min}]}(\norm{p_{contact}-p_{infected}}\le 6 feet)$, where $t_1$ and $t_2$ denote the time of illness onset of the infected person and the time the patient is isolated, respectively; $p_{infected}$ and $p_{contact}$ denote the positions of the infected person and the individual to be identified as a close contact, respectively.

In \cite{Xu2021PLOS}, we provided the first systematic control synthesis approach for three control strategies against COVID-19 with MTL specifications expressing the requirements for the control outcomes. In this paper, we extend \cite{Xu2021PLOS} in the following main aspects. (1) We investigate the \textit{robust} pandemic control synthesis problem that takes into account initial state and parameter uncertainties (which were not considered in \cite{Xu2021PLOS}). (2) For the pandemic control models that we consider, we derive the theoretical bounds for the \textit{robustness degree} of any trajectory within the initial state and parameter uncertainty ranges with respect to an MTL specification. (3) Based on the derived theoretical bounds, we develop an iterative optimization-based approach and in each iteration we solve a mixed-integer bi-linear programming problem (for vaccination control) or mixed-integer fractional constrained programming problem (for \textit{shield immunity} control \cite{Weitz_nature}).

% We prove that for the pandemic control models that we consider, the \textit{robustness degree} of any trajectory within the initial state and parameter uncertainty ranges with respect to any MTL specification can be theoretically bounded using the robustness degree of the \textit{nominal} trajectory and another variable that depends on the control inputs. 

% for two control strategies against a pandemic. We use \textit{metric temporal logic} (MTL) formulas to specify the expected control outcomes. The proposed control synthesis approach is based on two specific pandemic mitigation models: the susceptible, exposed, infectious, recovered (SEIR) model with vaccination control; and the SEIR model with shield immunity control. We develop methods for synthesizing control strategies based on the two specific pandemic models with MTL specifications. 

% Specifically, we convert the synthesis problem into mixed-integer bi-linear programming or mixed-integer fractional constrained programming problems, and solve the optimization problems using highly efficient solvers \cite{beal2018gekko}.

We provide simulation results in two different scenarios for robust control of the COVID-19 pandemic: one for vaccination control, and another for shield immunity control, with the model parameters estimated from data in Lombardy, Italy. The results show that the proposed synthesis approach can generate control inputs such that the time-varying numbers of individuals in each category (e.g., infectious, immune) satisfy the MTL specifications with robustness against initial state and parameter uncertainties.

\section{Related Work}

\noindent\textbf{Modeling and optimal control of pandemics}: There have been numerous research focusing on modeling the infection of pandemic diseases. Among the various models, \textit{compartmental models} such as the susceptible, infectious, and recovered (SIR) model \cite{Cooper2020} and its variations \cite{reiner2020modeling,Carcione_model,Feng2020,SIQS2021,Zhao2020,Giordano2020} have been commonly used. More detailed models have also been proposed to incorporate the hospitalized population \cite{Ivorra2020} and differentiate symptomatic and asymptomatic infected populations \cite{Giordano2020,Zhao2020Staggered,Mwalili2020}. There exist work in optimal control based on compartmental models \cite{Alonso2012,Pinho2015}. On the other hand, \textit{agent-based models} have been increasingly used by researchers to examine complex urban health problems and has been recently applied to study pandemics  \cite{Hoertel2020,cuevas2020agent,silva2020covid,inoue2020propagation,chang2020modelling}. While these models consider the details on the agent level, it is extremely difficult to identify optimal control policies for a large geographic region using agent-based models. There has also been work on analyzing or predicting the spread of pandemic diseases using artificial intelligence models \cite{Zheng_AI2020,Liu2020}, stochastic branching process \cite{althouse2020stochasticity}, etc. However, how to utilize such models in the optimal pandemic control synthesis is still an open problem. 

% \noindent\textbf{Optimal control of pandemic models}: 

\noindent\textbf{Formal control synthesis with temporal logic specifications}: The control synthesis approaches with temporal logic specifications mainly convert the control synthesis problem into a mixed-integer linear programming (MILP) problem~\cite{BluSTL,Allerton2019,zheACC2019DF,sayan2016,zhe_advisory,zheACC2018wind} which can be solved efficiently by MILP solvers. Another set of approaches substitute the temporal logic constraint into the objective function of the optimization problem and apply a functional gradient descent algorithm on the resulting unconstrained problem \cite{Andygradient,Abbas2014,zhe_control,zheACCstorageControl}. The control synthesis approach in this paper essentially extended the MILP-based approaches to non-linear dynamical systems to accommodate the pandemic models.

\section{Metric Temporal Logic, Trajectories, and Interval Trajectories}        
\label{sec_MTL}   
In this section, we briefly review  metric temporal logic (MTL)~\cite{FainekosMTL}. The state $x$ (e.g., representing the susceptible, exposed, infectious, recovered population of a certain region) belongs to
the domain $\mathcal{X}\subset\mathbb{R}^n_{\ge0}$. The time set is $\mathbb{T} = \mathbb{R}_{\ge0}$. The domain $\mathbb{B} = \{\textrm{True}, \textrm{False}\}$ is the Boolean domain, and the time index set is $\mathbb{I} = \{0,1,\dots\}$. We use $\xi: \mathbb{T}\mapsto\mathcal{X}$ to denote a \textit{trajectory}. We use $t[k]\in\mathbb{T}$ to denote the time instant at time index $k\in\mathbb{I}$ and $\xi_k\triangleq x(t[k])$ to denote the value of $x$ at time $t[k]$. A set $AP$ is a set of atomic propositions, each mapping from $\mathcal{X}$ to $\mathbb{B}$. The syntax of MTL is defined recursively as follows:
\[
\phi:=\top\mid \pi\mid\lnot\phi\mid\phi_{1}\wedge\phi_{2}\mid\phi_{1}\vee
\phi_{2}\mid\phi_{1}\mathcal{U}_{\mathcal{I}}\phi_{2},
\]
where $\top$ stands for the Boolean constant True, $\pi\in AP$ is an atomic
proposition, $\lnot$ (negation), $\wedge$ (conjunction), $\vee$ (disjunction)
are standard Boolean connectives, $\mathcal{U}$ is a temporal operator
representing \textquotedblleft until\textquotedblright, $\mathcal{I}$ is a time index interval of
the form $\mathcal{I}=[i_{1},i_{2}]$ ($i_1\le i_2$, $i_1, i_2\in\mathbb{I}$). In the remaining of this paper, we will use \textbf{day} as the unit in $\mathcal{I}$. We
can also derive two useful temporal operators from \textquotedblleft
until\textquotedblright~($\mathcal{U}$), which are \textquotedblleft
eventually\textquotedblright~$\Diamond_{\mathcal{I}}\phi=\top\mathcal{U}_{\mathcal{I}}\phi$ and
\textquotedblleft always\textquotedblright~$\Box_{\mathcal{I}}\phi=\lnot\Diamond_{\mathcal{I}}\lnot\phi$. 

We define the set of states that satisfy the atomic proposition $\pi$ as $\mathcal{O}(\pi)\subset \mathcal{X}$. We denote $\langle\langle\phi\rangle\rangle(\xi,k)=\top$ if the state of the trajectory $\xi$ satisfies the formula $\phi$ at time
instant $t[k]$ ($k\in\mathbb{I}$). Then the Boolean semantics of MTL are defined recursively as follows~\cite{FAINEKOScontinous}:
	\[   
	\begin{split}
	\langle\langle\top\rangle\rangle(\xi,k) :=& \top,\\
	\langle\langle \pi\rangle\rangle(\xi,k)  :=& \big(\xi_k\in\mathcal{O}(\pi)\big),\\
	\langle\langle \neg\phi\rangle\rangle(\xi,k)  :=&\neg\langle\langle \phi\rangle\rangle(\xi,k),\\
	\langle\langle\phi_1\vee\phi_2\rangle\rangle(\xi,k):=&\langle\langle\phi_1\rangle\rangle(\xi,k)\vee\langle\langle\phi_2\rangle\rangle(\xi,k),\\
	\langle\langle\phi_1\mathcal{U}_{\mathcal{I}}\phi_{2}\rangle\rangle(\xi,k)  :=&\bigvee_{k'\in (k+\mathcal{I})}\big(\langle\langle \phi_2\rangle\rangle(\xi,k')\wedge\bigwedge_{k\le k''<k'}\langle\langle \phi_1\rangle\rangle\\
	&(\xi,k'')\big), 
	\end{split}
	\]
	where $k+\mathcal{I}=\{k+\tilde{k}\vert \tilde{k}\in\mathcal{I}\}$.

	We denote the distance from $x\in\mathcal{X}$ to a set $S\subseteq\mathcal{X}$ as \textbf{dist}$_d(x,S)\triangleq$inf$\{d(x, y)\vert y\in cl(S)\}$ where $d$ is a metric on $\mathcal{X}$ and $cl(S)$ denotes the closure of the set $S$. In this paper, we use the metric $d(x,y)=\norm{x-y}_{\infty}$, where $\left\Vert\cdot\right\Vert_{\infty}$ denotes the infinity norm. We denote the depth of $x$ in $S$ as \textbf{depth}$_d(x,S)\triangleq$ \textbf{dist}$_d(x,\mathcal{X}\setminus S)$, the signed distance from $x$ to $S$ as
	\begin{equation}
	\textbf{Dist$_d(x,S)\triangleq$}%
	\begin{cases}
	-\textbf{dist}_d(x,S)& \mbox{if $x$ $\not\in S$};\\  
	\textbf{depth}_d(x,S) & \mbox{if $x$ $\in S$}.
	\end{cases}                        
	\end{equation}
	
	We use $\rho(\xi, \phi, k)$ to denote the robustness degree of the trajectory $\xi$ with respect to the formula $\phi$ at discrete-time
    instant $t[k]$ ($k\in\mathbb{I}$). The robust semantics of a formula $\phi$ with respect to $\xi$ are defined recursively as follows~\cite{FAINEKOScontinous}:
	\begin{align}
	\begin{split}
	\rho(\xi, \top, k) :=& +\infty,\\
	 \rho(\xi, \pi, k)  :=&\textbf{Dist$_d(\xi_k,\mathcal{O}(\pi))$},\\
	 \rho(\xi, \neg\phi, k)  :=&- \rho(\xi, \phi,k),\\
	\rho(\xi, \phi_1\vee\phi_2, k)  :=&\max\big( \rho(\xi, \phi_1, k), \rho(\xi, \phi_2, k)\big),\\
	\rho(\xi, \phi_1\mathcal{U}_{\mathcal{I}}\phi_{2}, k)  :=&\max_{k'\in (k+\mathcal{I})}\Big(\min\big( \rho(\xi, \phi_2, k'),\\& \min_{k\le k''<k'}
	\rho(\xi,\phi_1,k'')\big)\Big). 
	\end{split}
	\label{robustness_degree}
	\end{align}	
% The Boolean semantics of MTL can be found in \cite{FAINEKOScontinous}, with the slight variation that we only evaluate the satisfaction of a trajectory with respect to an MTL formula at discrete-time instants $t[k]~(k\in\mathcal{I})$. The robustness degree of a trajectory $\xi$ with respect to an MTL formula $\phi$ at time index $k$, denoted as $\phi(\xi, k)$, is defined recursively as follows:
% \[
% \begin{split}
% \top(\xi, k) :=& +\infty,\\
%  \pi(\xi, k)  :=&\textbf{Dist$_d(\xi_k,\mathcal{O}(\pi))$},\\
%  \neg\phi(\xi, k)  :=&- \phi(\xi, k),\\
% \phi_1\vee\phi_2(\xi, k)  :=&\max\big( \phi_1(\xi, k), \phi_2(\xi, k)\big),\\
% \phi_1\mathcal{U}_{\mathcal{I}}\phi_{2}(\xi, k)  :=&\max_{k'\in (k+\mathcal{I})}\Big(\min\big( \phi_2(\xi, k'),\\& \min_{k\leq k''<k'}\phi_1
% (\xi,k'')\big)\Big).  
% \end{split} 
% \]
% As defined, $\phi(\xi, k)\ge0$ if $\xi$ satisfies $\phi$ at time index $k$. 

% \section{Metric Temporal Logic with Interval Trajectories}

We denote an \textit{interval} by $[\underline{c}, \overline{c}] = \{ x \in \mathbb{R}^n | \underline{c}^i \leq x^i \leq \overline{c}^i, i=1,\dots, n \}$ for some $\underline{c}, \overline{c}\in \mathbb{R}^n$ such that $\underline{c}^i \leq \overline{c}^i$ holds for any $i\in\{1,\dots, n\}$ (we use the superscript $i$ to denote the $i$-th dimension). We define an \textit{interval trajectory}, denoted by $[\underline{\xi},\overline{\xi}]$, as a set of trajectories such that for any trajectory $\xi\in[\underline{\xi},\overline{\xi}]$ we have $\xi_k\in [\underline{\xi}_k, \overline{\xi}_k]$ holds for all $k$. For an interval trajectory $[\underline{\xi},\overline{\xi}]$, we define its \textit{nominal trajectory} $\xi^{\ast}$ such that $\xi^{\ast}_k=\frac{\underline{\xi}_k+\overline{\xi}_k}{2}$ for all $k$.

We use $\rho([\underline{\xi},\overline{\xi}], \phi, k)$ to denote the robustness degree of an interval trajectory $[\underline{\xi},\overline{\xi}]$ with respect to the formula $\phi$ at time instant $t[k]$ ($k\in\mathbb{I}$), and $\rho([\underline{\xi},\overline{\xi}], \phi, k)$ is defined as 
	\begin{align}
	\begin{split}
	\rho([\underline{\xi},\overline{\xi}], \phi, k) :=& \min_{\xi\in[\underline{\xi},\overline{\xi}]}\rho(\xi, \phi, k).
	\end{split}
	\label{max_RD}
	\end{align}	
Intuitively, if $\rho([\underline{\xi},\overline{\xi}], \phi, k)\ge0$, then any trajectory $\xi\in[\underline{\xi},\overline{\xi}]$ satisfied the MTL formula $
\phi$ at time instant $t[k]$.

% The robustness degree of $[\underline{\xi},\overline{\xi}]$ with respect to formula $\phi$ also satisfy the followings:
% 	\[
% 	\begin{split}
% 	\rho([\underline{\xi},\overline{\xi}], \top, k) =& +\infty,\\
% 	 \rho([\underline{\xi},\overline{\xi}], \pi, k)  =&\min_{\xi_k\in [\underline{\xi}_k, \overline{\xi}_k]}\textbf{Dist$_d(\xi_k,\mathcal{O}(\pi))$},\\
% 	\rho([\underline{\xi},\overline{\xi}], \phi_1\vee\phi_2, k)  =&\max\big( \rho([\underline{\xi},\overline{\xi}], \phi_1, k), \rho([\underline{\xi},\overline{\xi}], \phi_2, k)\big),\\
% 	\rho([\underline{\xi},\overline{\xi}], \phi_1\mathcal{U}_{\mathcal{I}}\phi_{2}, k)  =&\max_{k'\in (k+\mathcal{I})}\Big(\min\big( \rho([\underline{\xi},\overline{\xi}], \phi_2, k'),\\& \min_{k\lek''<k'}
% 	\rho([\underline{\xi},\overline{\xi}],\phi_1,k'')\big)\Big). 
% 	\end{split}
% 	\]	
	
% We note that $\rho([\underline{\xi},\overline{\xi}], \neg\phi, k)  =&- \rho([\underline{\xi},\overline{\xi}], \phi,k)$ generally does not hold.

\section{Pandemic SEIR Model with Control Strategies}
\label{sec_models}
In this section, we study the susceptible, exposed, infectious, recovered (SEIR) model for pandemics  \cite{Carcione_model,Weitz_nature,Zhao2020} with vaccination control and shield immunity control, and provide the problem formulation for robust pandemic control with formal specifications.

\begin{figure}
	\centering
	\includegraphics[scale=0.2]{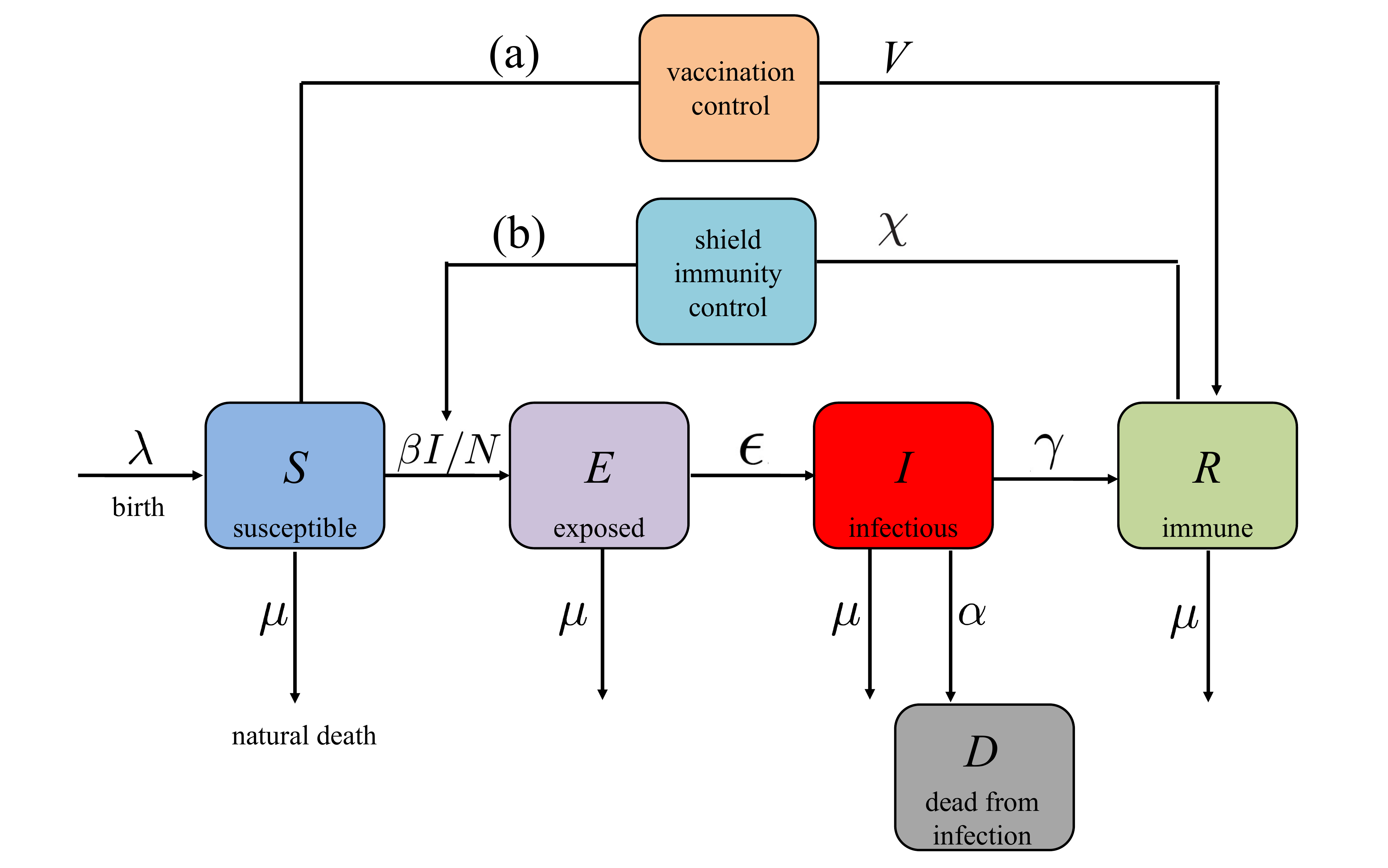}
	\caption{Block diagram of the pandemic SEIR model with (a) vaccination control and (b) shield immunity control.}
	\label{diagram_SEIR}
\end{figure}

As shown in Figure \ref{diagram_SEIR}, the total population is divided into five parts in an SEIR model: 
\begin{itemize}
    \item The susceptible population $S$: everyone is susceptible to the disease by birth since immunity is not hereditary; 
    \item The exposed population $E$: the individuals who have been exposed to the disease, but are still not infectious;
    \item The infectious population $I$: the individuals who are infectious;   
    \item The immune (recovered) population $R$: the individuals who are vaccinated or recovered from the disease, i.e., the population who are immune to the disease;
    \item The dead population $D$: the deaths from the disease.
\end{itemize}

\noindent\textbf{Vaccination control model:} We consider the SEIR model \cite{Carcione_model,Elie2020} with vaccination control as follows. 
\begin{align}
\begin{split}
&\dot{I} = \epsilon E - (\gamma+\mu+\alpha)I;\\
&\dot{E} = \beta SI/N - (\mu+\epsilon)E;\\
&\dot{S} = \lambda N - \mu S - \beta SI/N - V;\\
&\dot{R} = \gamma I - \mu R + V;\\
&\dot{D} = - \dot{I}-\dot{E}-\dot{S}-\dot{R},
\end{split}            
\label{vaccination_model}
\end{align}
where the control input $V$ is the number of vaccinated individuals per day, $N = S + E + I + R\le N_0$ is the total population in the region ($N_0$ is the initial total population in the region), $S$, $E$, $I$, and $R$ are the number of susceptible, exposed, infectious and recovered population in the region, respectively, and $D$ is the number of deaths from the pandemic disease in the region. For the parameters, $\lambda$ denotes the per-capita birth rate, 
$\mu$ is the per-capita natural death rate (death rate from causes unrelated to the pandemic disease), $\alpha$ is the pandemic virus-induced average fatality rate, 
$\beta$ is the probability of disease transmission per contact (dimensionless) times the number of contacts per unit
time, $\epsilon$ is the rate of progression from exposed to infectious (the
reciprocal is the incubation period), and $\gamma$ is the recovery rate of infectious individuals. We assume that the birth rate and the natural death rate are the same for the population we are
investigating, i.e., $\lambda=\mu$, and as a result, $D = N_0-I-E-S-R=N_0-N$ holds.

% Note that in (\ref{vaccination_model}), $D = N_0-I - E - S - R=N_0-N$ holds as we have assumed that the birth rate and the natural death rate are the same for the population we are investigating, i.e., $\lambda=\mu$.

\noindent\textbf{Shield immunity control model:} We consider the SEIR model with shield immunity control \cite{Weitz_nature} as follows (see Fig. \ref{diagram_SEIR} (b) as an illustration).
\begin{align}
\begin{split}
& \dot{I} = \epsilon E - (\gamma+\mu+\alpha)I ;\\
& \dot{E} = \beta SI/(N+\chi R) - (\mu+\epsilon)E;\\
& \dot{S} = \lambda N - \mu S - \beta SI/(N+\chi R);\\
& \dot{R} = \gamma I - \mu R;\\
& \dot{D} = - \dot{I}-\dot{E}-\dot{S}-\dot{R},
\end{split}      
\label{shield_model}
\end{align}
where the states and parameters are the same as in (\ref{vaccination_model}), while $\chi$ is the \textit{shield strength} \cite{Weitz_nature} as control input to be synthesized for the recovered population to substitute the contact for the susceptible population.

We rewrite (\ref{vaccination_model}) and (\ref{shield_model}) in the following general form.
\begin{align}
	\begin{split}
	&\dot{x}= f(x, u, \theta),
	\end{split} 
	\label{control_model}
\end{align}	
where the state $x = [I, E, S, R, D]^T\in\mathbb{R}^5_{\ge 0}$, the control input $u$ represents $V$ for the vaccination control and represents $\chi$ for the shield immunity control, $\theta= [\alpha, \beta, \epsilon, \gamma, \mu, \lambda]$, and $f:\mathbb{R}^5_{\ge 0}\times\mathbb{R}_{\ge 0}\times\mathbb{R}^6_{\ge 0}\rightarrow\mathbb{R}^5_{\ge 0}$ is a smooth vector field according to (\ref{vaccination_model}) and (\ref{shield_model}). 

For computational efficiency, we discretize the dynamics in (\ref{control_model}) as follows.
\begin{align}
	\begin{split}
	& x(k+1)= \bar{f}(x[k], u[k], \theta),
	\end{split} 
	\label{control_model2}
\end{align}	
where $\bar{f}(\bm\cdot, \bm\cdot, \bm\cdot)$ is discretized from $f(\bm\cdot, \bm\cdot, \bm\cdot)$ in (\ref{control_model}) using Euler's method.

Now we provide the problem formulation of the robust pandemic control synthesis problem as follows.
\begin{problem} [Robust pandemic control]
	Given the SEIR control model in (\ref{control_model2}) and an MTL specification $\phi$, compute the control input signal $u[\bm\cdot]$ that minimizes the control effort $\norm{u[\bm\cdot]}$ (here $\norm{\bm\cdot}$ denotes the $\ell_2$ norm), while guaranteeing $\rho([\underline{\xi},\overline{\xi}],\phi, 0) \ge 0$, where $[\underline{\xi},\overline{\xi}]$ is the interval trajectory starting with $x[0]\in[\underline{x}[0], \overline{x}[0]]$ with the control input signal $u[\bm\cdot]$ and parameter $\theta\in[\underline{\theta}, \overline{\theta}]$. 
	\label{problem1}                               
\end{problem}

\section{Solution} 
\label{sec_control}

The robust pandemic control synthesis problem can be formulated as a robust optimization problem as follows.
\begin{align}
\begin{split}
\underset{u[\bm\cdot]}{\min} ~ & \norm{u[\bm\cdot]}   \\
\text{s.t.} ~
& x[k+1] = \bar{f}(x[k], u[k], \theta), \forall k=0,\dots,T,  \\
& ~~~~~~~~~~\forall x[0]\in [\underline{x}[0], \overline{x}[0]], \forall \theta\in[\underline{\theta}, \overline{\theta}],\\
& 0\le u[k]\le u_{\textrm{max}}, \forall k=0,\dots,T,  \\ 
& \rho([\underline{\xi},\overline{\xi}],\phi, 0) \ge 0, 
\end{split}
\label{robust_formulaton}
\end{align} 
where $[\underline{\xi},\overline{\xi}]$ is the interval trajectory starting with $x[0]\in[\underline{x}[0], \overline{x}[0]]$ with the control input signal $u[\bm\cdot]$ and parameter $\theta\in[\underline{\theta}, \overline{\theta}]$, and $T\in\mathbb{I}$ is the maximal time index we consider.

Generally, the optimization problem in (\ref{robust_formulaton}) is a robust mixed-integer non-linear programming problem. We refer the readers to \cite{sayan2016} for a detailed description of how the constraint $\rho(\xi,\phi, 0) \ge 0$ is encoded to satisfy an MTL specification $\phi$. The integer variables are introduced when a \textit{big-M formulation} \cite{Schrijver86} is needed to satisfy MTL specifications that contain $\Diamond_{\mathcal{I}}$ or $\vee$. 

To efficiently solve the robust optimization problem in (\ref{robust_formulaton}), we first provide the following theorem.
\begin{theorem} 
	\label{gamma}
Given an interval trajectory $[\underline{\xi},\overline{\xi}]$ and its nominal trajectory $\xi^{\ast}$, then
for any $\xi\in [\underline{\xi},\overline{\xi}]$ and any $k\in \{0,1,\dots, K\}$, we have
	\begin{align} \nonumber
	&\rho(\xi^{\ast},\phi, k)-\delta_{\rm{max}}\le \rho(\xi,\phi, k)\le \rho(\xi^{\ast},\phi, k)+\delta_{\rm{max}},
	\end{align}  
where $\phi$ is any MTL formula, $\delta_{\rm{max}}\triangleq\max\limits_{k}\delta_k$, and $\delta_k=\max\limits_{i}\Big(\frac{\overline{\xi}^i_k-\underline{\xi}^i_k}{2}\Big)$, and $\overline{\xi}^i_k$ is the $i$-th dimension value of $\overline{\xi}_k$.                     	                       
\end{theorem}

From Theorem \ref{gamma}, it can be seen that if we can design the control input signal such as $\rho(\xi^{\ast},\phi, k)\ge\delta_{\rm{max}}$, then we have $\rho(\xi,\phi, k)\ge0$ holds for any $\xi\in[\underline{\xi},\overline{\xi}]$, i.e.,
$\rho([\underline{\xi},\overline{\xi}], \phi, k)\ge0$. However, as $\delta_k=\max\limits_{i}\Big(\frac{\overline{\xi}^i_k-\underline{\xi}^i_k}{2}\Big)$ depends on $u$, $\delta_{\rm{max}}$ also depends on $u$. Therefore, we need to design an iterative approach 
to compute $u$ such that $\rho(\xi^{\ast},\phi, k)\ge\delta_{\rm{max}}$. Such an iterative approach is shown in Algorithm \ref{robust_alg}.

\begin{algorithm}[h!]
	\caption{Robust pandemic control synthesis with MTL specifications.}                                                               
	\label{MTLalg}
	\begin{algorithmic}[1]
		\State \textbf{Inputs:}  $[\underline{x}[0],\overline{x}[0]]$, $[\underline{\theta}, \overline{\theta}]$, $\bar{f}$
		\State Initialize $\zeta\gets0$, $u$
		\State Compute interval trajectory $[\underline{\xi},\overline{\xi}]$ with control input signal $u[\bm\cdot]$ for $x[0]\in [\underline{x}[0], \overline{x}[0]]$, $\theta\in[\underline{\theta}, \overline{\theta}]$	
		\State Compute $\rho([\underline{\xi},\overline{\xi}], \phi, 0)$ and $\delta_{\rm{max}}$
		\State $Iter\gets1$
        \While{$(\rho([\underline{\xi},\overline{\xi}], \phi, 0)<0) \wedge (Iter<Iter_{\textrm{max}})$}
        
        \State Solve (\ref{nominal_formulaton}) to obtain the optimal control input signal $~~~~~~u^{\ast}[\bm\cdot]$ with robustness $\delta_{\rm{max}}$
        \If{(\ref{nominal_formulaton}) is infeasible}
        \State $\zeta\gets\zeta-\rho([\underline{\xi},\overline{\xi}], \phi, 0)$
        \State Solve (\ref{nominal_formulaton}) to obtain the optimal control inputs $u^{\ast}$ $~~~~~~~~$ with robustness $\zeta$ (i.e., replace $\delta_{\rm{max}}$ with $\zeta$ in (\ref{nominal_formulaton}))
        \EndIf
        \State $u[\bm\cdot]\gets u^{\ast}[\bm\cdot]$
       \State Compute interval trajectory $[\underline{\xi},\overline{\xi}]$ with $u$ for $x[0]\in$ $~~~~~~[\underline{x}[0], \overline{x}[0]]$, $\theta\in[\underline{\theta}, \overline{\theta}]$	
       \State Compute $\rho([\underline{\xi},\overline{\xi}], \phi, 0)$ and $\delta_{\rm{max}}$
       \State $Iter\gets Iter+1$
		\EndWhile    		                
		\State Return $u$
	\end{algorithmic}
	\label{robust_alg}
\end{algorithm}

% \begin{algorithm}[h!]
% 	\caption{Robust control synthesis with MTL specifications.}                                                                   
% 	\label{MTLalg}
% 	\begin{algorithmic}[1]
% 		\State \textbf{Inputs:}  $[\underline{x}[0],\overline{x}[0]]$
% 		\State Initialize $\gamma$, $u$
% 		\State Simulate interval trajectory $[\underline{\xi},\overline{\xi}]$ with control inputs $u$ and compute robustness degree $\rho([\underline{\xi},\overline{\xi}], \phi, 0)$
%         \While{$\rho([\underline{\xi},\overline{\xi}], \phi, 0)<0$}
%         \State $\gamma\gets\gamma-\rho([\underline{\xi},\overline{\xi}], \phi, 0)$
%         \State Compute the optimal control inputs $u^{\ast}$ with robustness $\gamma$
%         \State $u\gets u^{\ast}$
%       \State Simulate interval trajectory $[\underline{\xi},\overline{\xi}]$ with control inputs $u$ and compute robustness degree $\rho([\underline{\xi},\overline{\xi}], \phi, 0)$        
% 		\EndWhile    		                
% 		\State Return $u$
% 	\end{algorithmic}
% 	\label{robust_alg}
% \end{algorithm}

In each iteration in the while loop, we solve the following optimization problem for synthesizing the control input signal $u[\bm\cdot]$ such that the nominal trajectory satisfies the MTL specification $\phi$ with robustness of at least $\delta_{\rm{max}}$.
\begin{align}
\begin{split}
\underset{u[\bm\cdot]}{\min} ~ & \norm{u[\bm\cdot]}   \\
\text{s.t.} ~
& x^{\ast}[k+1] = \bar{f}(x^{\ast}[k], u[k], \theta^{\ast}), \forall k=0,\dots,T,  \\
& 0\le u[k]\le u_{\textrm{max}}, \forall k=0,\dots,T,  \\ 
& \rho(\xi^{\ast},\phi, 0) \ge \delta_{\rm{max}}. 
\end{split}
\label{nominal_formulaton}
\end{align} 

In (\ref{nominal_formulaton}), we approximate the total population $N$ with the initial population $N_0$ as the change of $N$ is relatively small compared to the multiplication of the susceptible population and the infectious population. With such an approximation, the optimization problem becomes a mixed-integer bi-linear programming problem (for vaccination control) or mixed-integer fractional constrained programming problem (for shield immunity control), which can be efficiently solved through solvers such as GEKKO \cite{beal2018gekko}.

% which can be efficiently solved through techniques such as McCormick's relaxation \cite{McCormick1976,Gupte2013SolvingMI}. 

If the constrained optimization problem in (\ref{nominal_formulaton}) is infeasible (e.g., due to the conservativeness of the bound $\delta_{\rm{max}}$), we will re-solve (\ref{nominal_formulaton}) by replacing $\delta_{\rm{max}}$ with $\zeta-\rho([\underline{\xi},\overline{\xi}], \phi, 0)$ (where $\zeta$ is initially set as 0). After we obtain the optimal control input signal $u^{\ast}[\bm\cdot]$ from solving (\ref{nominal_formulaton}), we compute interval trajectory $[\underline{\xi},\overline{\xi}]$ with $u^{\ast}$ for $x[0]\in [\underline{x}[0], \overline{x}[0]]$, $\theta\in[\underline{\theta}, \overline{\theta}]$. Then we compute $\rho([\underline{\xi},\overline{\xi}], \phi, 0)$ based on (\ref{robustness_degree}) and (\ref{max_RD}). If $\rho([\underline{\xi},\overline{\xi}], \phi, 0)\ge0$, then the algorithm terminates with the solution $u^{\ast}$; otherwise, we update $\zeta$ as $\zeta-\rho([\underline{\xi},\overline{\xi}], \phi, 0)$ and repeat the above procedures until either $\rho([\underline{\xi},\overline{\xi}], \phi, 0)\ge0$ holds or a maximal number of iterations (denoted as $Iter_{\textrm{max}}$) is reached.

% & \underline{x}[0]\le x[0]\le \overline{x}[0], \nonumber \\ 
% & \underline{\theta}\le \theta\le \overline{\theta}, \nonumber \\ 

% The above optimization problem is generally a mixed-integer non-linear programming problem. We refer the readers to \cite{sayan2016} for a detailed description of how the constraint $\langle\langle\varphi_{\textrm{V}}\rangle\rangle(\xi^{\textrm{V}}_{\bm\cdot;x^{init}_{\textrm{V}},V}, 0)=\top$ is encoded to satisfy an MTL specification $\varphi_{\textrm{V}}$. The integer variables are introduced when a \textit{big-M formulation} \cite{Schrijver86} is needed to satisfy MTL specifications such as $\Diamond_{[0,10]}\varphi$ ($\varphi$ should hold true for at least one day during the first 10 days) or $\varphi_1\vee\varphi_2$ (at least one of the MTL formulas $\varphi_1$, $\varphi_2$ should hold true). As the change of total population is relatively small compared to the multiplication of the susceptible population and the infectious population, we approximate the term $T_{\textrm{s}}\beta S[k]I[k]/N[k]$ with $T_{\textrm{s}}\beta S[k]I[k]/N_0$. With such an approximation, the optimization problem becomes a mixed-integer bi-linear programming problem, which can be more efficiently solved using techniques such as McCormick's relaxation \cite{McCormick1976,Gupte2013SolvingMI}. Furthermore, if the MTL specification $\varphi$ consists of only conjunctions ($\wedge$) and the always operator ($\Box$), the integers in the optimization problem can be eliminated \cite{sayan2016} and the problem becomes a bi-linear programming problem.

\section{Simulation Results} 
In this section, we implement the proposed robust control synthesis methods in the COVID-19 models estimated from data in Lombardy, Italy.

\subsection{Robust Vaccination Control for COVID-19}
\label{results_vaccination}
The parameters of the COVID-19 SEIR model with uncertainties are shown in Table \ref{parameter_SEIR}. They were estimated in \cite{Carcione_model} from the data in the early days (from February 23 to March 16, 2020) in Lombardy, Italy with no isolation measures. The start time for the simulations in this subsection is February 23, 2020. We consider three MTL specifications as shown in Table \ref{result_vaccination}. For example, $\phi_{\textrm{V}}^1=\Box_{[0,100]} (I\le 0.3)\wedge\Box_{[0,100]} (D\le0.05)\wedge\Diamond_{[40, 60]}(R\ge8)$, which means ``the infected population should never exceed 0.3 million and the deceased population should never exceed 0.05 million within the next 100 days, and the immune population should eventually exceed 8 million after 40 to 60 days''. We choose the initial values of the states with uncertainties as $I[0]=1000\pm1000$ (people), $E[0]=0.02\pm0.001$ million, $S[0]=9.979\pm0.001$ million, $R[0]=0$ and $D[0]=0$. 

% Fig. \ref{fig_results0} shows the simulation results without any vaccination. It can be seen that the three MTL specifications $\phi_{\textrm{V}}^1$, $\phi_{\textrm{V}}^2$ and $\phi_{\textrm{V}}^3$ are all violated in such a situation. Note that as isolation measures (i.e., home isolation, social distancing and partial national lockdown) were taken since March 16 in Lombardy, Italy, the real situation was better than those shown in Fig. \ref{fig_results0}. Now we investigate the hypothetical scenario where the isolation measures are replaced by vaccination.

	\begin{table}[]
		\centering
		\caption{Parameters of COVID-19 SEIR model estimated from data from Lombardy, Italy from February 23 to March 16 (2020) with no isolation measures \cite{Carcione_model} with uncertainties.}
		\label{parameter_SEIR}  
		\begin{tabular}{llll}
			\toprule[2pt]    
			parameter    & value & parameter    & value\\ \hline
		    ~~~~$\epsilon$        & 0.2$\pm$0.001/day & ~~~~$\lambda$        & 1/30295 \\ 
		   ~~~~$\gamma$        & 0.2$\pm$0.001/day &~~~~$\mu$        & 1/30295   \\ 
		    ~~~~$\alpha$        & 0.006$\pm$0.001/day &  ~~~~$N_0$         & 10 million \\ 
		    ~~~~$\beta$       & 0.75$\pm$0.001/day & ~~~~$T_{\textrm{s}}$  & 1 day
            \\ \bottomrule[2pt]       
		\end{tabular}
	\end{table}    
	
		\begin{table}[]
		\centering
		\caption{MTL specifications and simulation results for vaccination control.}
		\label{result_vaccination}  
		\begin{tabular}{ll>{\raggedright\arraybackslash}p{23mm}}
		\toprule[2pt]    
			 ~~~~~MTL specification & \tabincell{c}{control\\ effort} & \tabincell{c}{computation time\\(each iteration)}\\ \hline
		    \tabincell{c}{$\phi_{\textrm{V}}^1=\Box_{[0,100]} (I\le0.3)$\\ $~~~~~~\wedge\Box_{[0,100]} (D\le0.05)$\\ $~~~\wedge\Diamond_{[40, 60]}(R\ge8)$}   & ~1.28 & ~~~1.365 s  \\ 
		     \tabincell{c}{$\phi_{\textrm{V}}^2=\Box_{[0,100]} (I\le0.15)$\\ $~~~~~\wedge\Box_{[0,100]} (D\le0.02)$\\
		     $~~\wedge\Diamond_{[40, 60]}(R\ge9)$}  & ~2.397 & ~~~1.134 s \\ 
		     \tabincell{c}{$\phi_{\textrm{V}}^3=\Box_{[0,100]} (I\le0.1)$\\ $~~~~~\wedge\Box_{[0,100]} (D\le0.01)$\\$\wedge\Diamond_{[40, 60]}(R\ge9)$} & ~6.934 & ~~~3.289 s \\ \bottomrule[2pt]       
		\end{tabular}
	\end{table}

	 \begin{table}[!h]
		\centering
		\caption{MTL specifications and simulation results for shield immunity control.}
		\begin{tabular}{ll>{\raggedright\arraybackslash}p{23mm}}
		\toprule[2pt]    
			 ~~~~MTL specification & \tabincell{c}{control\\ effort} & \tabincell{c}{computation time\\
			 (each iteration)}\\ \hline
		    \tabincell{c}{$\varphi_{\textrm{S}}^1=\Box_{[0,100]} (I\le0.6)$\\ $~~~~~~\wedge\Box_{[0,100]} (D\le0.1)$\\ $~~~\wedge\Diamond_{[40, 60]}(R\ge1)$}   & 33349.80 & ~~~3.498 s  \\ 
		     \tabincell{c}{$\varphi_{\textrm{S}}^2=\Box_{[0,100]} (I\le0.5)$\\ $~~~~~\wedge\Box_{[0,100]} (D\le0.07)$\\
		     $~~\wedge\Diamond_{[40, 60]}(R\ge1)$}  & 84272.22 & ~~~3.312 s \\ 
		     \tabincell{c}{$\varphi_{\textrm{S}}^3=\Box_{[0,100]} (I\le0.3)$\\ $~~~~~\wedge\Box_{[0,100]} (D\le0.06)$\\$\wedge\Diamond_{[40, 60]}(R\ge1)$} & 122476.59 & ~~~2.385 s \\ \bottomrule[2pt]       
		\end{tabular}
		\label{result_shield}  
	\end{table} 

\begin{figure*}
	\centering
	\includegraphics[scale=0.25]{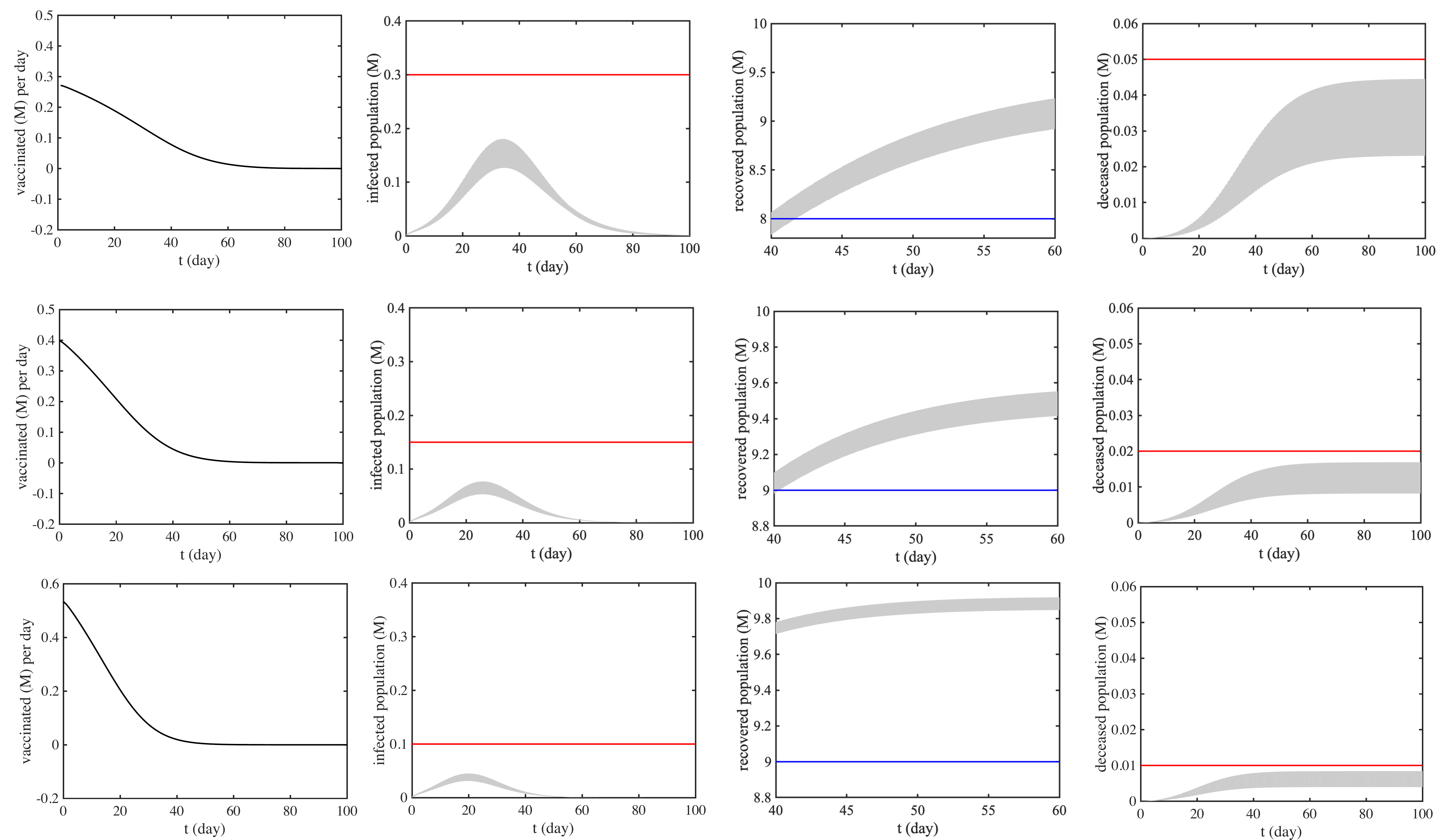}
	\caption{Optimal vaccination control input signals and interval trajectories (plotted with shading) for COVID-19 SEIR model with robust vaccination control and MTL specifications $\varphi_{\textrm{V}}^1$ (first row), $\varphi_{\textrm{V}}^2$ (second row) and $\varphi_{\textrm{V}}^3$ (third row). The red and blue lines indicate the thresholds that should never be exceeded and should eventually be exceeded in the atomic propositions of the MTL specifications, respectively.}
	\label{fig_results}
\end{figure*}

We used the the CORA toolbox \cite{CORA} to compute $[\underline{\xi}, \overline{\xi}]$ with the initial state and parameter uncertainties. We use the solver GEKKO \cite{beal2018gekko} to solve the optimization problems formulated in Section \ref{sec_control}. We set $Iter_{\textrm{max}}=100$, while in reality the algorithm terminates in all cases within three iterations with feasible and optimal solutions. Fig. \ref{fig_results} and Table \ref{result_vaccination} show the simulation results for vaccination control of COVID-19 SEIR model with MTL specifications $\phi_{\textrm{V}}^1$, $\phi_{\textrm{V}}^2$ and $\phi_{\textrm{V}}^3$, respectively. The results show that, despite the initial state and parameter uncertainties, the MTL specifications $\phi_{\textrm{V}}^1$, $\phi_{\textrm{V}}^2$ and $\phi_{\textrm{V}}^3$ are satisfied with the synthesized vaccination control input signals respectively. It can be seen that vaccination within the first 40 days after the outbreak can mitigate the spread of COVID-19 in the most efficient manner. The results also show that more control efforts are needed for more stringent specifications. 

% For all three specifications, the computations are completed within 4 seconds on a MacBook Laptop with 2.80-GHz Core i9 CPU and 32-GB RAM.

% \begin{figure}
% 	\centering
% 	\begin{subfigure}[b]{0.22\textwidth}
% 		\centering
% 		\includegraphics[width=\textwidth]{case0_a.png}
% 		\caption{Number of individuals}
% 	\end{subfigure} 
% 		\hfill
% 	\begin{subfigure}[b]{0.22\textwidth}
% 		\centering
% 		\includegraphics[width=\textwidth]{case0_c.png}
% 		\caption{Number of deaths}
% 	\end{subfigure}
% 			\hfill
% 	\begin{subfigure}[b]{0.22\textwidth}
% 		\centering
% 		\includegraphics[width=\textwidth]{case0_d.png}
% 		\caption{Number of deaths per day}
% 	\end{subfigure}
% 	\caption{Simulation results for COVID-19 SEIR model estimated from data from Lombardy, Italy with no isolation measures.}  
% 	\label{fig_results0}
% \end{figure}

\begin{figure*}
	\centering
	\includegraphics[scale=0.25]{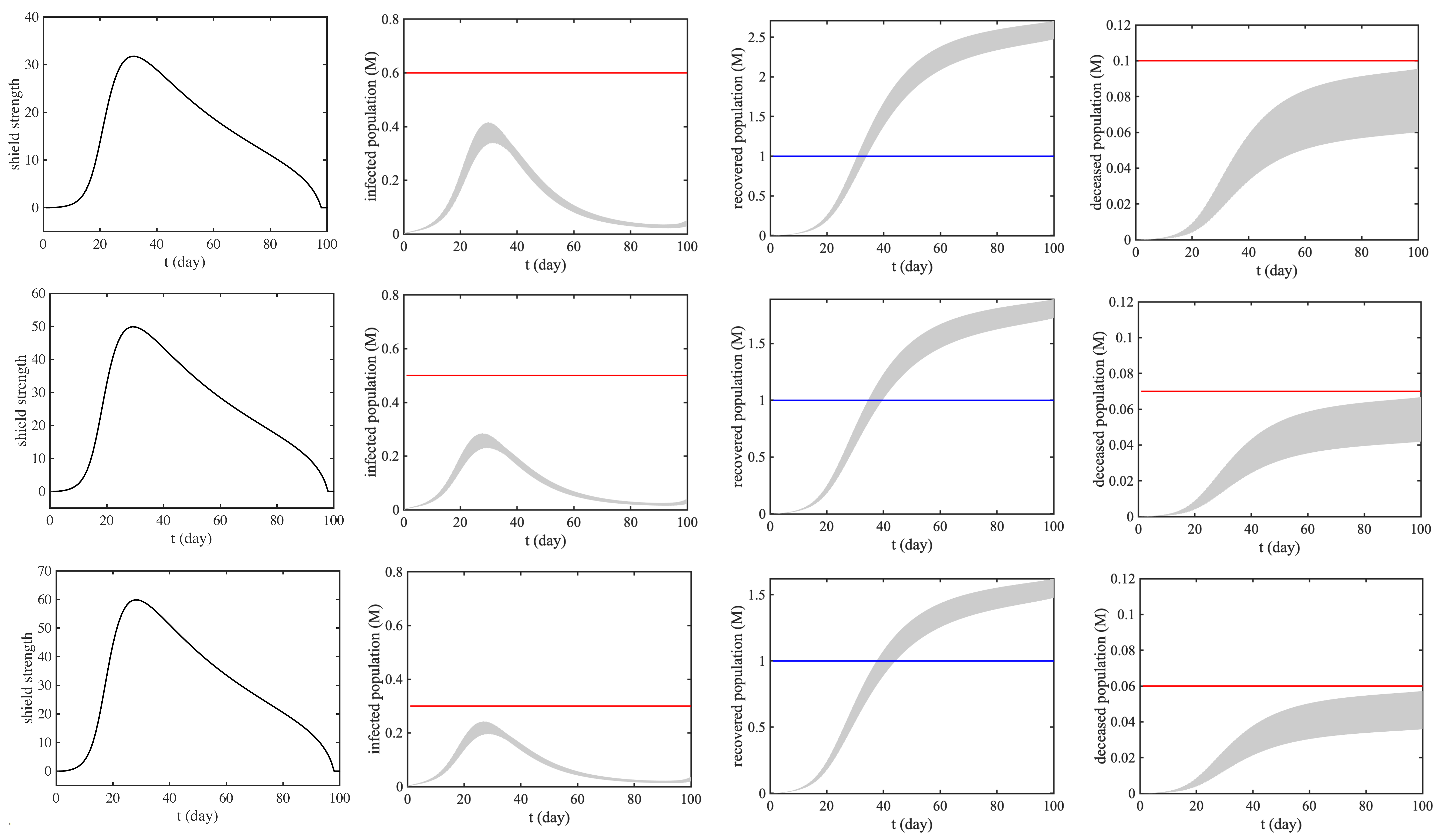}
	\caption{Optimal shield immunity control input signals and interval trajectories (plotted with shading) for COVID-19 SEIR model with robust shield immunity control and MTL specifications $\varphi_{\textrm{S}}^1$ (first row), $\varphi_{\textrm{S}}^2$ (second row) and $\varphi_{\textrm{S}}^3$ (third row). The red and blue lines indicate the thresholds that should never be exceeded and should eventually be exceeded in the atomic propositions of the MTL specifications, respectively.}
	\label{fig_results_shield}
\end{figure*}

\subsection{Robust Shield Immunity Control for COVID-19}
\label{results_shield_immunity}
We use the same initial state and parameter values of the COVID-19 SEIR model with uncertainties as shown in Table \ref{parameter_SEIR}. The start time for the simulations in this subsection is February 23, 2020. We set the three MTL specifications $\varphi_{\textrm{S}}^1$, $\varphi_{\textrm{S}}^2$ and $\varphi_{\textrm{S}}^3$ (as shown in Table \ref{result_shield}) to be less stringent than the MTL specifications with the vaccination control, as shield immunity is generally less effective than vaccination. We investigate the hypothetical scenario where the isolation measures are replaced by shield immunity control.

Fig. \ref{fig_results_shield} and Table \ref{result_shield} show the simulation results for shield immunity control of the COVID-19 SEIR model with MTL specifications $\varphi_{\textrm{S}}^1$, $\varphi_{\textrm{S}}^2$ and $\varphi_{\textrm{S}}^3$, respectively. The results show that, despite the initial state and parameter uncertainties, the MTL specifications $\varphi_{\textrm{S}}^1$, $\varphi_{\textrm{S}}^2$ and $\varphi_{\textrm{S}}^3$ are satisfied respectively. We observe that with the three MTL specifications, the synthesized shield immunity control input signals all increase to a peak after approximately 20 to 40 days and then gradually decrease. These observations indicate that shield immunity at early days of COVID-19 is more efficient than shield immunity at later days. The results also show that more control efforts are needed for more stringent specifications. 

\section{Conclusion}
In this paper, we proposed a systematic control synthesis approach for mitigating a pandemic based on two control models with vaccination and shield immunity, respectively. The proposed approach can synthesize control inputs that lead to satisfaction of metric temporal logic specifications despite the state and parameter uncertainties. 

We list two future directions as follows. First, we will extend this work to online control synthesis so that the states and parameters can be updated periodically with the latest disease infection data. Second, we will study the benefits and costs of joint control of different control strategies so that the specifications can be satisfied with coordinated efforts.

\section*{Appendix}
\textbf{Proof of Theorem \ref{gamma}}:\\
To prove Theorem \ref{gamma}, we first prove that Theorem \ref{gamma} holds for any atomic proposition $\pi$.   

As the metric $d$ satisfies the triangle inequality, for any $k$, we have that for any $\xi_k\in [\underline{\xi}_k,\overline{\xi}_k]$ and any $y\in\mathcal{X}$,
\begin{align}
\begin{split}
&  d(\xi^{\ast}_{k},y)-d(\xi^{\ast}_{k},\xi_k) \le d(\xi_{k},y)
\le d(\xi^{\ast}_{k},y)+d(\xi^{\ast}_{k},\xi_k).
\end{split}          
\label{tri}                                                         
\end{align}       

As $d(\xi^{\ast}_{k},\xi_k)\le\max\limits_{i}\Big(\frac{\overline{\xi}^i_k-\underline{\xi}^i_k}{2}\Big)=\delta_k$, we have
\begin{align}
\begin{split}
&  d(\xi^{\ast}_{k},y)-\delta_k \le d(\xi_{k},y)
\le d(\xi^{\ast}_{k},y)+\delta_k.
\end{split}          
\label{tri}                                                         
\end{align}   

% In the following, we denote $\mathcal{Z}([\underline{\xi}_k, \overline{\xi}_k])\triangleq\{\xi_{k}~\vert~ \xi_k\in [\underline{\xi}_k, \overline{\xi}_k]\}$.

\begin{figure}
	\centering
	\includegraphics[scale=0.2]{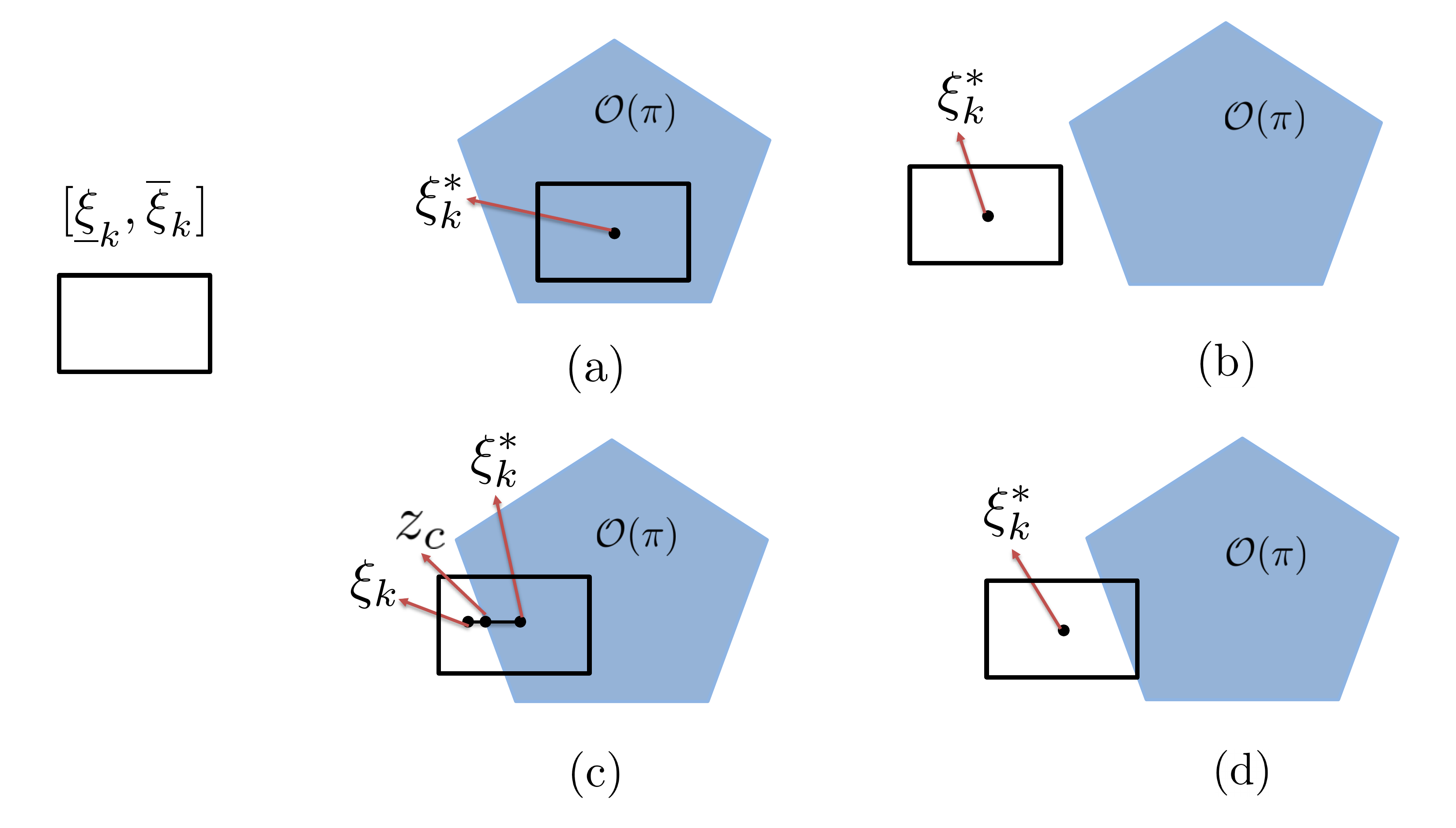}
	\caption{Four different cases in the proof: (a) $\xi^{\ast}_{k}\in \mathcal{O}(\pi)$,  $[\underline{\xi}_k, \overline{\xi}_k]\subset\mathcal{O}(\pi)$; (b) $\xi^{\ast}_{k}\notin \mathcal{O}(\pi)$,  $[\underline{\xi}_k, \overline{\xi}_k]\subset\mathcal{X}\backslash\mathcal{O}(\pi)$; (c) $\xi^{\ast}_{k}\in \mathcal{O}(\pi)$,  $[\underline{\xi}_k, \overline{\xi}_k]\not\subset\mathcal{O}(\pi)$; (d) $\xi^{\ast}_{k}\notin \mathcal{O}(\pi)$,  $[\underline{\xi}_k, \overline{\xi}_k]\not\subset\mathcal{X}\backslash\mathcal{O}(\pi)$.} 
	\label{proof}
\end{figure}

1) $\xi^{\ast}_{k}\in \mathcal{O}(\pi)$, and $[\underline{\xi}_k, \overline{\xi}_k]\subset\mathcal{O}(\pi)$, as shown in Fig. \ref{proof} (a). In this case, for any $\xi_{k}\in [\underline{\xi}_k, \overline{\xi}_k]$, 
\begin{align}\nonumber
&\rho(\xi,\pi, k)=\mbox{inf}\{d(\xi_{k},y)|y \in \mathcal{X}\backslash\mathcal{O}(\pi)\}.                                     
\end{align}
Thus from (\ref{tri}), we have
\begin{align}\nonumber
&\rho(\xi,\pi, k)\ge\mbox{inf}\{d(\xi^{\ast}_{k},y)-\delta_{k}|y \in \mathcal{X}\backslash\mathcal{O}(\pi)\}\\\nonumber
& =\mbox{inf}\{d(\xi^{\ast}_{k},y)|y \in \mathcal{X}\backslash\mathcal{O}(\pi)\}-\delta_{k}  \\\nonumber  & =\rho(\xi^{\ast},\pi, k)-\delta_{k}.                               
\end{align}			

2) $\xi^{\ast}_{k}\notin \mathcal{O}(\pi)$, and $[\underline{\xi}_k, \overline{\xi}_k]\subset\mathcal{X}\backslash\mathcal{O}(\pi)$, as shown in Fig. \ref{proof} (b). In this case, for any $\xi_{k}\in [\underline{\xi}_k, \overline{\xi}_k]$,
\begin{align}\nonumber
&\rho(\xi,\pi, k)=-\mbox{inf}\{d(\xi_{k},y)|y \in cl(\mathcal{O}(\pi))\}.                                     
\end{align}
Thus from (\ref{tri}), we have
\begin{align}\nonumber
&\rho(\xi,\pi, k)\ge-\mbox{inf}\{d(\xi^{\ast}_{k},y)+\delta_{k}|y \in cl(\mathcal{O}(\pi))\}\\\nonumber
& =\rho(\xi^{\ast},\pi, k)-\delta_{k}.                                    
\end{align}		

3) $\xi^{\ast}_{k}\in \mathcal{O}(\pi)$, but $[\underline{\xi}_k, \overline{\xi}_k]\not\subset\mathcal{O}(\pi)$, as shown in Fig. \ref{proof} (c). In this case, we have
\begin{align} \nonumber
\begin{split}
&\rho(\xi,\pi, k)\ge\min\limits_{\xi_{k}\in [\underline{\xi}_k, \overline{\xi}_k]}\rho(\xi,\pi, k)=\min\{X_1, X_2\}, \\
& \mbox{where}\\
&X_1=-\max_{\substack{\xi_{k}\in [\underline{\xi}_k, \overline{\xi}_k],\\\xi_{k}\notin \mathcal{O}(\pi)}}\mbox{inf}\{d(\xi_{k},y)|y \in cl(\mathcal{O}(\pi))\},\\
&X_2=\min_{\substack{\xi_{k}\in [\underline{\xi}_k, \overline{\xi}_k],\\\xi_{k}\in \mathcal{O}(\pi)}}\mbox{inf}\{d(\xi_{k},y)|y \in \mathcal{X}\backslash\mathcal{O}(\pi)\}.
\end{split}
\end{align}      

As $d(\xi_{k},y)\ge0$, so $X_1\le0, X_2\ge0$, $\min\{X_1, X_2\}=X_1$. For any $\xi_{k}\in [\underline{\xi}_k, \overline{\xi}_k]$ and $\xi_{k}\notin \mathcal{O}(\pi)$, there exists $z_c\in [\underline{\xi}_k, \overline{\xi}_k]$ and $z_c\in \partial(\mathcal{O}(\pi))$ such that $\xi_{k}, z_c$ and $\xi^{\ast}_{k}$ are collinear, i.e.
\begin{align}\nonumber
&d(\xi^{\ast}_{k},z_c)+d(z_c,\xi_{k})=d(\xi^{\ast}_{k},\xi_{k})\le\delta_{k}.
\end{align}		
Therefore, as 
$\rho(\xi^{\ast},\pi, k)=\mbox{inf}\{d(\xi^{\ast}_{k},y)|y \in \mathcal{X}\backslash\mathcal{O}(\pi)\}\le d(\xi^{\ast}_{k},z_c)$ and $\mbox{inf}\{d(\xi_{k},y)|y \in cl(\mathcal{O}(\pi))\}\le d(\xi_{k},z_c)$, we have for any $x\in B(\hat{x}^{0},\delta_{k})$ and $\xi_{k}\notin \mathcal{O}(\pi)$,
\begin{align}\nonumber
\mbox{inf}\{d(\xi_{k},y)|y \in cl(\mathcal{O}(\pi))\}+&\rho(\xi^{\ast},\pi, k)\le\delta_{k}.                                     
\end{align}

So $-X_1+\rho(\xi^{\ast},\pi, k)\le\delta_{k}$, i.e. $X_1\ge\rho(\xi^{\ast},\pi, k)-\delta_{k}$. Therefore, $\rho(\xi,\pi, k)\ge\min\{X_1, X_2\}=X_1\ge\rho(\xi^{\ast},\pi, k)-\delta_{k}$.

4) $\xi^{\ast}_{k}\notin \mathcal{O}(\pi)$, but  $[\underline{\xi}_k, \overline{\xi}_k]\not\subset\mathcal{X}\backslash\mathcal{O}(\pi)$, as shown in Fig. \ref{proof} (d). In this case,
\begin{align}\nonumber
&\rho(\xi^{\ast},\pi, k)=-\mbox{inf}\{d(\xi^{\ast}_{k},y)|y \in cl(\mathcal{O}(\pi))\}.                                     
\end{align}		  
For any $\xi_{k}\in [\underline{\xi}_k, \overline{\xi}_k]$ and $\xi_{k}\notin\mathcal{O}(\pi)$, 
\begin{align}\nonumber
&\rho(\xi,\pi, k)=-\mbox{inf}\{d(\xi_{k},y)|y \in cl(\mathcal{O}(\pi))\}\\\nonumber
& \ge-\mbox{inf}\{d(\xi^{\ast}_{k},y)+\delta_{k}|y \in cl(\mathcal{O}(\pi))\}  \\\nonumber
& =\rho(\xi^{\ast},\pi, k)-\delta_{k}.                                 
\end{align}			
Therefore, $\rho(\xi,\pi, k)\ge\min\{X_1, X_2\}=X_1\ge\rho(\xi^{\ast},\pi, k)-\delta_{k}$.

In sum, we have proven that $\rho(\xi,\pi, k)\ge\rho(\xi^{\ast},\pi, k)-\delta_{k}$. Similarly, we can prove that $\rho(\xi,\pi, k)\le \rho(\xi^{\ast},\pi, k)+\delta_{k}$.

Therefore, Theorem \ref{gamma} holds for any atomic proposition $\pi$. Next, we use induction to prove that Theorem \ref{gamma} holds for any MTL formula $\phi$.

(ii) We assume that Theorem \ref{gamma} holds for $\phi$ and prove Theorem \ref{gamma} holds for $\lnot\phi$.  

If Theorem \ref{gamma} holds for $\phi$, then as $\rho(\xi^{\ast},\lnot\phi, k)=-\rho(\xi^{\ast},\phi, k)$, we have $-\rho(\xi^{\ast},\lnot\phi, k)-\delta_{\rm{max}}\le -\rho(\xi,\lnot\phi, k)\le -\rho(\xi^{\ast},\lnot\phi, k)+\delta_{\rm{max}}$, thus $\rho(\xi^{\ast},\lnot\phi, k)-\delta_{\rm{max}}\le \rho(\xi,\lnot\phi, k)\le \rho(\xi^{\ast},\lnot\phi, k)+\delta_{\rm{max}}$.

(iii) We assume that Theorem \ref{gamma} holds for $\phi_1,\phi_2$ and prove Theorem \ref{gamma} holds for $\phi_1\wedge\phi_2$. 

If Theorem \ref{gamma} holds for $\phi_1$ and $\phi_2$, then $\rho(\xi^{\ast},\phi_1, k)-\delta_{\rm{max}}\le \rho(\xi,\phi_1, k)\le \rho(\xi^{\ast},\phi_1, k)+\delta_{\rm{max}}$, $\rho(\xi^{\ast},\phi_2, k)-\delta_{\rm{max}}\le \rho(\xi,\phi_2, k)\le \rho(\xi^{\ast},\phi_2, k)+\delta_{\rm{max}}$. As $\rho(\xi^{\ast},\phi_1\wedge\phi_2, k)=\min(\rho(\xi^{\ast},\phi_1, k),\rho(\xi^{\ast},\phi_2, k))$, we have 
\begin{align}\nonumber
\begin{split}
&\min(\rho(\xi^{\ast},\phi_1, k),\rho(\xi^{\ast},\phi_2, k))-\delta_{\rm{max}}\\&\le\rho(\xi,\phi_1\wedge\phi_2, k) \\
&\le\min(\rho(\xi^{\ast},\phi_1, k), \rho(\xi^{\ast},\phi_2, k))+\delta_{\rm{max}},
\end{split}
\end{align}
therefore $\rho(\xi^{\ast},\phi_1\wedge\phi_2, k)-\delta_{\rm{max}}\le$ $\rho(\xi,\phi_1\wedge\phi_2, k)\le \rho(\xi^{\ast},\phi_1\wedge\phi_2, k)+\delta_{\rm{max}}$.                                            

(iv) We assume that Theorem \ref{gamma} holds for $\phi$ and prove Theorem \ref{gamma} holds for $\phi_1\mathcal{U}_{\mathcal{I}}\phi_2$.   

As
\begin{align}\nonumber 
\begin{split}
\rho(\xi^{\ast},\phi_1\mathcal{U}_{\mathcal{I}}\phi_2, k)=&\max\limits_{k'\in (t+\mathcal{I})}\Big(\min\big(\rho(\xi^{\ast},\phi_2, k'), \\& \min\limits_{t\le k''<k'}\left[\left[\phi_1\right]\right]
(\xi^{\ast},k'')\big)\Big), 
\end{split}
\end{align}
if Theorem \ref{gamma} holds for $\phi_1$ and $\phi_2$, then $\rho(\xi^{\ast},\phi_1, k'')-\delta_{\rm{max}}\le \rho(\xi,\phi_1, k'')\le \rho(\xi^{\ast},\phi_1, k'')+\delta_{\rm{max}}$, $\rho(\xi^{\ast},\phi_2, k')-\delta_{\rm{max}}\le \rho(\xi,\phi_2, k')\le \rho(\xi^{\ast},\phi_2, k')+\delta_{\rm{max}}$, so we have 
\begin{align}\nonumber            
\begin{split}
&\max_{k'\in (t+\mathcal{I})}\Big(\min\big(\rho(\xi^{\ast},\phi_2, k'), \\
&~~~~~~~~~~~~~~\min_{t\le k''<k'}\rho
(\xi^{\ast},\phi_1,k'')\big)\Big)-\delta_{\rm{max}}\\
&\le\max_{k'\in (t+\mathcal{I})}\Big(\min\big(\rho(\xi,\phi_2, k'), \min_{t\le k''<k'}\rho
(\xi^{\ast},\phi_1,k'')\big)\Big)\\
&\le\max_{k'\in (t+\mathcal{I})}\Big(\min\big(\rho(\xi^{\ast},\phi_2, k'), \\
&~~~~~~~~~~~~~~\min_{t\le k''<k'}\rho
(\xi^{\ast},\phi_1,k'')\big)\Big)+\delta_{\rm{max}}.   
\end{split}
\end{align}
Thus Theorem \ref{gamma} holds for $\phi_1\mathcal{U}_{\mathcal{I}}\phi_{2}$.

Therefore, it is proved by induction that Theorem \ref{gamma} holds for any MTL formula $\phi$.

\bibliographystyle{IEEEtran}
\bibliography{references}

\begin{thebibliography}{10}
\providecommand{\url}[1]{#1}
\csname url@samestyle\endcsname
\providecommand{\newblock}{\relax}
\providecommand{\bibinfo}[2]{#2}
\providecommand{\BIBentrySTDinterwordspacing}{\spaceskip=0pt\relax}
\providecommand{\BIBentryALTinterwordstretchfactor}{4}
\providecommand{\BIBentryALTinterwordspacing}{\spaceskip=\fontdimen2\font plus
\BIBentryALTinterwordstretchfactor\fontdimen3\font minus
  \fontdimen4\font\relax}
\providecommand{\BIBforeignlanguage}[2]{{%
\expandafter\ifx\csname l@#1\endcsname\relax
\typeout{** WARNING: IEEEtran.bst: No hyphenation pattern has been}%
\typeout{** loaded for the language `#1'. Using the pattern for}%
\typeout{** the default language instead.}%
\else
\language=\csname l@#1\endcsname
\fi
#2}}
\providecommand{\BIBdecl}{\relax}
\BIBdecl

\bibitem{zhe_ijcai2019}
Z.~Xu and U.~Topcu, ``Transfer of temporal logic formulas in reinforcement
  learning,'' in \emph{Proc. IJCAI'2019}, 7 2019, pp. 4010--4018.

\bibitem{Verginis2019Icra}
C.~K. {Verginis}, C.~{Vrohidis}, C.~P. {Bechlioulis}, K.~J. {Kyriakopoulos},
  and D.~V. {Dimarogonas}, ``Reconfigurable motion planning and control in
  obstacle cluttered environments under timed temporal tasks,'' in \emph{2019
  International Conference on Robotics and Automation (ICRA)}, May 2019, pp.
  951--957.

\bibitem{zhe_control}
Z.~Xu, A.~Julius, and J.~H. Chow, ``Energy storage controller synthesis for
  power systems with temporal logic specifications,'' \emph{IEEE Systems
  Journal}, Early access on IEEE Xplore.

\bibitem{Xu2021PLOS}
\BIBentryALTinterwordspacing
Z.~Xu, B.~Wu, and U.~Topcu, ``Control strategies for {COVID-19} epidemic with
  vaccination, shield immunity and quarantine: A metric temporal logic
  approach,'' \emph{PLOS ONE}, vol.~16, no.~3, pp. 1--20, 03 2021. [Online].
  Available: \url{https://doi.org/10.1371/journal.pone.0247660}
\BIBentrySTDinterwordspacing

\bibitem{Weitz_nature}
J.~Weitz, S.~Beckett, A.~Coenen, D.~David, M.~Dominguez-Mirazo, J.~Dushoff,
  J.~Leung, G.~Li, A.~Magalie, S.~Park, R.~Rodriguez-Gonzalez, S.~Shivam, and
  C.~Zhao, ``Intervention serology and interaction substitution: Modeling the
  role of `shield immunity' in reducing {COVID-19} epidemic spread,''
  \emph{Nature Medicine}, pp. 849--854, 2020.

\bibitem{Cooper2020}
\BIBentryALTinterwordspacing
I.~Cooper, A.~Mondal, and C.~G. Antonopoulos, ``\BIBforeignlanguage{eng}{A sir
  model assumption for the spread of {COVID-19} in different communities},''
  \emph{\BIBforeignlanguage{eng}{Chaos, solitons, and fractals}}, vol. 139, pp.
  110\,057--110\,057, Oct 2020, 32834610[pmid]. [Online]. Available:
  \url{https://pubmed.ncbi.nlm.nih.gov/32834610}
\BIBentrySTDinterwordspacing

\bibitem{reiner2020modeling}
R.~C. Reiner, R.~M. Barber, J.~K. Collins, P.~Zheng, C.~Adolph, J.~Albright,
  C.~M. Antony, A.~Y. Aravkin, S.~D. Bachmeier, B.~Bang-Jensen \emph{et~al.},
  ``Modeling {COVID-19} scenarios for the united states,'' \emph{Nature
  medicine}, 2020.

\bibitem{Carcione_model}
\BIBentryALTinterwordspacing
J.~M. Carcione, J.~E. Santos, C.~Bagaini, and J.~Ba, ``A simulation of a
  {COVID-19} epidemic based on a deterministic {SEIR} model,'' \emph{Frontiers
  in Public Health}, vol.~8, p. 230, 2020. [Online]. Available:
  \url{https://www.frontiersin.org/article/10.3389/fpubh.2020.00230}
\BIBentrySTDinterwordspacing

\bibitem{Feng2020}
\BIBentryALTinterwordspacing
Z.~Feng, J.~W. Glasser, and A.~N. Hill, ``On the benefits of flattening the
  curve: A perspective,'' \emph{Mathematical Biosciences}, vol. 326, p. 108389,
  2020. [Online]. Available:
  \url{http://www.sciencedirect.com/science/article/pii/S0025556420300729}
\BIBentrySTDinterwordspacing

\bibitem{SIQS2021}
\BIBentryALTinterwordspacing
X.-B. Zhang and X.-H. Zhang, ``The threshold of a deterministic and a
  stochastic {SIQS} epidemic model with varying total population size,''
  \emph{Applied Mathematical Modelling}, vol.~91, pp. 749 -- 767, 2021.
  [Online]. Available:
  \url{http://www.sciencedirect.com/science/article/pii/S0307904X20305710}
\BIBentrySTDinterwordspacing

\bibitem{Zhao2020}
\BIBentryALTinterwordspacing
S.~Zhao and H.~Chen, ``Modeling the epidemic dynamics and control of {COVID-19}
  outbreak in {China},'' \emph{medRxiv}, 2020. [Online]. Available:
  \url{https://www.medrxiv.org/content/early/2020/03/09/2020.02.27.20028639}
\BIBentrySTDinterwordspacing

\bibitem{Giordano2020}
G.~Giordano, F.~Blanchini, R.~Bruno, P.~Colaneri, A.~{Di Filippo}, A.~{Di
  Matteo}, and M.~Colaneri, ``\BIBforeignlanguage{English}{Modelling the
  {COVID-19} epidemic and implementation of population-wide interventions in
  {Italy}},'' \emph{\BIBforeignlanguage{English}{Nature Medicine}}, Jan. 2020.

\bibitem{Ivorra2020}
B.~Ivorra, M.~Ruiz~Ferrández, M.~Vela, and A.~Ramos, ``Mathematical modeling
  of the spread of the coronavirus disease 2019 ({COVID-19}) taking into
  account the undetected infections. {The} case of {China},'' \emph{Commun
  Nonlinear Sci Numer Simul.}, April 2020.

\bibitem{Zhao2020Staggered}
``Staggered release policies for {COVID-19} control: Costs and benefits of
  relaxing restrictions by age and risk,'' \emph{Mathematical Biosciences},
  vol. 326, p. 108405, 2020.

\bibitem{Mwalili2020}
S.~Mwalili, M.~Kimathi, V.~Ojiambo, D.~Gathungu, and R.~Mbogo, ``{SEIR} model
  for {COVID-19} dynamics incorporating the environment and social
  distancing,'' \emph{BMC Research Notes}, vol.~13, no.~1, p. 352, 2020.

\bibitem{Alonso2012}
S.~Alonso-Quesada, M.~De~la Sen, R.~Agarwal, and A.~Ibeas, ``An observer-based
  vaccination control law for a {SEIR} epidemic model based on feedback
  linearization techniques for nonlinear systems,'' \emph{Advances in
  Difference Equations}, vol. 2012, 09 2012.

\bibitem{Pinho2015}
M.~d.~R. de~Pinho, I.~Kornienko, and H.~Maurer, ``Optimal control of a {SEIR}
  model with mixed constraints and {L1} cost,'' in \emph{CONTROLO'2014 --
  Proceedings of the 11th Portuguese Conference on Automatic Control}, A.~P.
  Moreira, A.~Matos, and G.~Veiga, Eds.\hskip 1em plus 0.5em minus 0.4em\relax
  Cham: Springer International Publishing, 2015, pp. 135--145.

\bibitem{Hoertel2020}
N.~Hoertel, M.~Blachier, M.~Olfson, M.~Massetti, M.~Rico, F.~Limosin, and
  H.~Leleu, ``A stochastic agent-based model of the {SARS-CoV-2} epidemic in
  france,'' \emph{Nature Medicine}, vol.~26, pp. 1--5, 09 2020.

\bibitem{cuevas2020agent}
E.~Cuevas, ``An agent-based model to evaluate the {COVID-19} transmission risks
  in facilities,'' \emph{Computers in Biology and Medicine}, p. 103827, 2020.

\bibitem{silva2020covid}
P.~C. Silva, P.~V. Batista, H.~S. Lima, M.~A. Alves, F.~G. Guimar{\~a}es, and
  R.~C. Silva, ``{COVID-ABS}: An agent-based model of {COVID-19} epidemic to
  simulate health and economic effects of social distancing interventions,''
  \emph{Chaos, Solitons \& Fractals}, vol. 139, p. 110088, 2020.

\bibitem{inoue2020propagation}
H.~Inoue and Y.~Todo, ``The propagation of the economic impact through supply
  chains: The case of a mega-city lockdown to contain the spread of
  {COVID-19},'' \emph{Covid Economics}, vol.~2, pp. 43--59, 2020.

\bibitem{chang2020modelling}
S.~L. Chang, N.~Harding, C.~Zachreson, O.~M. Cliff, and M.~Prokopenko,
  ``Modelling transmission and control of the {COVID-19} pandemic in
  australia,'' \emph{arXiv preprint arXiv:2003.10218}, 2020.

\bibitem{Zheng_AI2020}
N.~{Zheng}, S.~{Du}, J.~{Wang}, H.~{Zhang}, W.~{Cui}, Z.~{Kang}, T.~{Yang},
  B.~{Lou}, Y.~{Chi}, H.~{Long}, M.~{Ma}, Q.~{Yuan}, S.~{Zhang}, D.~{Zhang},
  F.~{Ye}, and J.~{Xin}, ``Predicting {COVID}-19 in {China} using hybrid {AI}
  model,'' \emph{IEEE Trans. Cybernetics}, vol.~50, no.~7, pp. 2891--2904,
  2020.

\bibitem{Liu2020}
\BIBentryALTinterwordspacing
F.~Liu, J.~Wang, J.~Liu, Y.~Li, D.~Liu, J.~Tong, Z.~Li, D.~Yu, Y.~Fan, X.~Bi,
  X.~Zhang, and S.~Mo, ``Predicting and analyzing the {COVID-19} epidemic in
  {China}: Based on {SEIRD}, {LSTM} and {GWR} models,'' \emph{PLOS ONE},
  vol.~15, no.~8, pp. 1--22, 08 2020. [Online]. Available:
  \url{https://doi.org/10.1371/journal.pone.0238280}
\BIBentrySTDinterwordspacing

\bibitem{althouse2020stochasticity}
B.~M. Althouse, E.~A. Wenger, J.~C. Miller, S.~V. Scarpino, A.~Allard,
  L.~H{\'e}bert-Dufresne, and H.~Hu, ``Stochasticity and heterogeneity in the
  transmission dynamics of {SARS-CoV-2},'' \emph{arXiv preprint
  arXiv:2005.13689}, 2020.

\bibitem{BluSTL}
A.~Donze and V.~Raman, ``{BluSTL}: Controller synthesis from signal temporal
  logic specifications,'' in \emph{ARCH14-15. 1st and 2nd International
  Workshop on Applied veRification for Continuous and Hybrid Systems}, ser.
  EPiC Series in Computing, G.~Frehse and M.~Althoff, Eds., vol.~34.\hskip 1em
  plus 0.5em minus 0.4em\relax EasyChair, 2015, pp. 160--168.

\bibitem{Allerton2019}
Z.~{Xu}, F.~M. {Zegers}, B.~{Wu}, W.~{Dixon}, and U.~{Topcu}, ``Controller
  synthesis for multi-agent systems with intermittent communication. a metric
  temporal logic approach,'' in \emph{Allerton'19}, pp. 1015--1022.

\bibitem{zheACC2019DF}
Z.~{Xu}, K.~{Yazdani}, M.~T. {Hale}, and U.~{Topcu}, ``Differentially private
  controller synthesis with metric temporal logic specifications,'' in
  \emph{2020 American Control Conference (ACC)}, 2020, pp. 4745--4750.

\bibitem{sayan2016}
S.~Saha and A.~A. Julius, ``An {MILP} approach for real-time optimal controller
  synthesis with metric temporal logic specifications,'' in \emph{Proc. IEEE
  Amer. Control Conf.}, July 2016, pp. 1105--1110.

\bibitem{zhe_advisory}
Z.~Xu, S.~Saha, B.~Hu, S.~Mishra, and A.~A. Julius, ``Advisory temporal logic
  inference and controller design for semiautonomous robots,'' \emph{IEEE
  Trans. Autom. Sci. Eng.}, pp. 1--19, 2018.

\bibitem{zheACC2018wind}
Z.~Xu, A.~A. Julius, and J.~H. Chow, ``Coordinated control of wind turbine
  generator and energy storage system for frequency regulation under temporal
  logic specifications,'' in \emph{Proc. Amer. Control Conf.}, 2018, pp.
  1580--1585.

\bibitem{Andygradient}
A.~K. Winn and A.~A. Julius, ``Optimization of human generated trajectories for
  safety controller synthesis,'' in \emph{Proc. IEEE Amer. Control Conf.}, June
  2013, pp. 4374--4379.

\bibitem{Abbas2014}
H.~Abbas, A.~Winn, G.~Fainekos, and A.~A. Julius, ``Functional gradient descent
  method for metric temporal logic specifications,'' in \emph{Proc. IEEE Amer.
  Control Conf.}, June 2014, pp. 2312--2317.

\bibitem{zheACCstorageControl}
Z.~Xu, A.~Julius, and J.~H. Chow, ``Optimal energy storage control for
  frequency regulation under temporal logic specifications,'' in \emph{2017
  American Control Conference (ACC)}, May 2017, pp. 1874--1879.

\bibitem{FainekosMTL}
G.~E. Fainekos and G.~J. Pappas\vspace{0mm}, ``Robustness of temporal logic
  specifications,'' in \emph{Formal Approaches to Testing and Runtime
  Verification, in: LNCS, vol. 4262, Springer, 2006}.

\bibitem{FAINEKOScontinous}
G.~E. Fainekos and G.~J. Pappas, ``Robustness of temporal logic specifications
  for continuous-time signals,'' \emph{Theoretical Computer Science}, vol. 410,
  no.~42, pp. 4262 -- 4291, 2009.

\bibitem{Elie2020}
R.~Elie, E.~Hubert, and G.~Turinici, ``Contact rate epidemic control of
  {COVID-19}: an equilibrium view,'' 04 2020.

\bibitem{Schrijver86}
A.~Schrijver, \emph{Theory of Linear and Integer Programming}.\hskip 1em plus
  0.5em minus 0.4em\relax John Wiley \& Sons, Chichester, 1986.

\bibitem{beal2018gekko}
L.~Beal, D.~Hill, R.~Martin, and J.~Hedengren, ``Gekko optimization suite,''
  \emph{Processes}, vol.~6, no.~8, p. 106, 2018.

\bibitem{CORA}
M.~Althoff, ``An introduction to cora 2015,'' in \emph{ARCH14-15. 1st and 2nd
  International Workshop on Applied veRification for Continuous and Hybrid
  Systems}, ser. EPiC Series in Computing, G.~Frehse and M.~Althoff, Eds.,
  vol.~34.\hskip 1em plus 0.5em minus 0.4em\relax EasyChair, 2015, pp.
  120--151.

\end{thebibliography}
	
\end{document}